\documentclass[prb,draft,amsfonts,amssymb,amsmath,showpacs]{revtex4}
\begin{document}

\title{Relativistic BCS Theory in Quasi-(2+1)-Dimensions: Effects of an Inter-layer Transfer in the Honeycomb Lattice Symmetry}
\author{Tadafumi Ohsaku}
\affiliation{Institute of Physics, Academy of Sciences of the Czech Republic, Na Slovance 2, CZ-18221 Prague, Czech Republic}
\email[Present Address; Department of Physics, University of Texas at Austin,]{tadafumi@physics.utexas.edu}
\date{\today}

\newcommand{\bmx}{\mbox{\boldmath $x$}}
\newcommand{\bmp}{\mbox{\boldmath $p$}}
\newcommand{\bmk}{\mbox{\boldmath $k$}}
\newcommand{\bmq}{\mbox{\boldmath $q$}}
\newcommand{\Afey}{\ooalign{\hfil/\hfil\crcr$A$}}
\newcommand{\kfey}{\ooalign{\hfil/\hfil\crcr$k$}}
\newcommand{\pfey}{\ooalign{\hfil/\hfil\crcr$p$}}
\newcommand{\partfey}{\ooalign{\hfil/\hfil\crcr$\partial$}}
\newcommand{\Dfey}{\ooalign{\hfil/\hfil\crcr$D$}}
\newcommand{\hfey}{\ooalign{\hfil/\hfil\crcr$h$}}

\begin{abstract}

Motivated by recent huge interests on graphene sheet and graphite as "relativistic" systems, 
a BCS superconductivity in a quasi-(2+1)-dimensional relativistic model is investigated.
The intra-layer particle dynamics is described by a (2+1)-dimensional Gross-Neveu-type
four-body contact interaction model, while inter-layer particle motions will be caused by a hopping term
with a transfer parameter $t$.
Especially, we examine the effects of non-vanishing $t$, chemical potential $\mu$,
and a mass parameter $m$ which will give a gap at a conical intersection point of the relativistic band dispersion,
in the BCS $s$-wave ( scalar ) superconducting gap function $\Delta$.

\end{abstract}


\maketitle

The honeycomb lattice structure has been obtained remarkable interests recently in condensed matter physics.
For example, graphite and a graphene sheet have the honeycomb lattice symmetry, 
and they show various interesting physical properties~[1-22].
It is well-known fact that, a band structure of a system of honeycomb symmetry has conical intersections between the valence
and conduction bands, and the band dispersions can be described by a Dirac model.
Graphite can be regarded as a system where (2+1)-dimensional layers are stacked in a specific direction.
Toward some efficient design for electronic substances, 
it is desirable for us to find simple but powerful tendencies/laws 
( especially as a function of inter-layer transfer $t$,
which is approximatedly given by an overlapping integral between orbitals of two layers ).
A study by employing a simple field-theoretical model seems suitable for us for this purpose.

\vspace{2mm}

Toward an examination on superconductivity in the honeycomb lattice symmetry, 
the author studied a BCS-type theory in a single layer system 
( the exact (2+1)-dimensional case ) in Ref.~[7] as a first step.
Recently, Ref.~[5] investigated a kind of BCS-type superconductivity in a lattice model of graphene sheet.
It is interesting for us to investigate some effects coming from an inter-layer interaction ( particle hopping between two nearest neighbor layers )
in relativistic BCS superconductivity. 
Of course, such motions of particles by hopping between layers surely exist in real substances,
except the case of an ideal mono-layer situation. 
Therefore, we will investigate a BCS theory in a quasi-(2+1)-dimensional system where $x$ and $y$ directions are homogeneous,
while particle motions of $z$-direction are described by a Hubbard-like hopping ( lattice ) model in this paper. 

\vspace{2mm}

The energy spectra of particles near a conical intersection point with having a hopping term 
with a transfer parameter $t$ perpendicular to two-dimensional layers will be given as follows:
\begin{eqnarray}
E^{(0)}_{\pm}(\bmk) &=& \pm |k_{\perp}| - t\cos k_{3}, \quad k_{\perp} \equiv (k_{1}, k_{2}).
\end{eqnarray}
We will seek a model Lagrangian which will give the energy spectra of (1) from its kinetic term, while it also has a BCS-type interaction between particles.
The candidate for us will be found in the following form:
\begin{eqnarray}
{\cal L} &\equiv&  {\cal L}_{K}  + {\cal L}_{I}, \\
{\cal L}_{K} &\equiv& 
 \bar{\psi}(x_{0},x_{\perp},n)\Bigl[\gamma^{0} ( i\partial_{0}+\mu ) + i\gamma^{\perp}\cdot \partial_{\perp} -m\Bigr]\psi(x_{0},x_{\perp},n)   \nonumber \\
& & 
+ \frac{1}{2}\bar{\psi}(x_{0},x_{\perp},n)\Bigl[ \gamma^{0}t  -u \Bigr] \psi(x_{0},x_{\perp},n+1)  
+ \frac{1}{2}\bar{\psi}(x_{0},x_{\perp},n+1) \Bigl[ \gamma^{0}t  -u \Bigr]  \psi(x_{0},x_{\perp},n),   \\
{\cal L}_{I} &\equiv& G\Bigl[\bar{\psi}(x_{0},x_{\perp},n)\psi(x_{0},x_{\perp},n)\Bigr]^{2},
\end{eqnarray}
( a similar model was discussed in Ref.~[23] ).
${\cal L}_{K}$ has both the usual Dirac kinetic term in (2+1)-dimensions and the inter-layer hopping terms.
The abbreviations for coordinates and derivatives inside the layer ( i.e., perpendicular to the third, stacking direction ) are defined by
\begin{eqnarray}
x_{\perp} \equiv (x_{1},x_{2}),  \quad   \partial_{\perp} \equiv (\partial_{1},\partial_{2}),
\end{eqnarray}
while, we choose the following definition of the three-demensional gamma matrices:
\begin{eqnarray}
\gamma^{0} \equiv \sigma_{3}, \quad \gamma^{1} \equiv i\sigma_{1}, \quad \gamma^{2} \equiv i\sigma_{2},  \quad
\gamma^{\perp} \equiv ( \gamma^{1},\gamma^{2}). 
\end{eqnarray}
$\psi$ is a two-component spinor.
The Fourier transform of the fermion field in our model becomes
\begin{eqnarray}
\psi(x_{0},x_{\perp},n) &=& \int_{k}\psi(k_{0},k_{\perp},k_{3})e^{-ik_{0}x_{0}+ik_{\perp}\cdot x_{\perp}+ik_{3}n},  \\
\int_{k}(\cdots) &\equiv& \int\frac{dk_{0}}{2\pi}\int\frac{d^{2}k_{\perp}}{(2\pi)^{2}}\int^{\pi}_{-\pi}\frac{dk_{3}}{2\pi}(\cdots).
\end{eqnarray} 
The integration of $k_{3}$ is restricted to the region $-\pi\le k_{3}\le \pi$, namely the first Brillouin zone.
Thus, the Fourier transformed expression for the kinetic Lagrangian is found to be
\begin{eqnarray}
{\cal L}_{K} &=& \bar{\psi}\Bigl(\gamma^{0}\{ k_{0} + \mu + t\cos k_{3} \} + \gamma^{\perp}\cdot k_{\perp} -m-u\cos k_{3}  \Bigr)\psi.
\end{eqnarray}
The diagonalization of ${\cal L}_{K}$ at $\mu=m=u=0$ gives (1) as its eigenvalues.
Our Lagrangian ${\cal L}$ becomes a (2+1)-dimensional Gross-Neveu model~[7,24,25] at the zero transfer limit $t\to 0$, $u\to 0$.
$u\cos k_{3}$ is a mass with an alternation ( modulation ) in the $k_{3}$-direction.

\vspace{2mm}

In this paper, we only consider scalar-type fermion pair condensates $\sigma$ ( fermion-antifermion ) and $\Delta$ ( fermion-fermion ).
From possible functional ( mathematical ) structure of gap equations, this choice is relevant for considering BCS superconductivity~[7].
We introduce the method of auxiliary fields of composites to our Lagrangian:
\begin{eqnarray}
{\cal L} &=&  {\cal L}_{K}  -\sigma\bar{\psi}\psi + \frac{\Delta}{2}\bar{\psi}C\bar{\psi}^{T} + \frac{\overline{\Delta}}{2}\psi^{T}C^{-1}\psi
-\frac{\sigma^{2}}{2G} -\frac{|\Delta|^{2}}{2G}     \nonumber \\
&=& -\frac{1}{2G}\Bigl( \sigma^{2} + |\Delta|^{2} \Bigr) + \frac{1}{2}\bar{\Psi}{\cal M}\Psi.
\end{eqnarray}
Here, $C$ is the charge conjugation matrix, 
and we have taken the four-component Nambu notation~[26] of the following definition:
\begin{eqnarray}
\Psi(x) \equiv \left(
\begin{array}{c}
\psi(x) \\
\bar{\psi}^{T}(x)
\end{array}
\right), \quad \bar{\Psi}(x) = (\bar{\psi}(x),\psi^{T}(x)),  \quad 
{\cal M}  \equiv
\left(
\begin{array}{cc}
{\cal M}_{11} & {\cal M}_{12} \\
{\cal M}_{21} & {\cal M}_{22}
\end{array}
\right),  
\end{eqnarray}
where,
\begin{eqnarray}
& & {\cal M}_{11} \equiv \gamma^{0}\{ i\partial_{0} +\mu+t\cos k_{3} \} +i\gamma^{\perp}\cdot \partial_{\perp} -\sigma -m -u\cos k_{3},  \nonumber \\
& & {\cal M}_{12} \equiv \Delta(x) C, \quad 
{\cal M}_{21} \equiv \overline{\Delta}(x)C^{-1},  \nonumber  \\
& & {\cal M}_{22} \equiv \gamma^{0T}\{ i\partial_{0} - \mu-t\cos k_{3} \} +i(\gamma^{\perp})^{T}\cdot \partial_{\perp} +\sigma + m + u\cos k_{3}.
\end{eqnarray}
Hence,
the partition function ${\cal Z}$ considered in this paper is defined as follows:
\begin{eqnarray}
{\cal Z} &=& \int {\cal D}\psi{\cal D}\bar{\psi}{\cal D}\sigma{\cal D}\Delta{\cal D}\overline{\Delta}\exp\Bigl[ i\sum_{n}\int dx_{0}dx_{1}dx_{2} {\cal L}\Bigr]. 
\end{eqnarray}
The effective potential becomes
\begin{eqnarray}
V_{eff} &=&  \frac{\sigma^{2}}{2G} + \frac{|\Delta|^{2}}{2G}   
+ i {\rm tr} \int_{k}   \ln {\rm det}{\cal M},  
\end{eqnarray}
where the determinant is evaluated to be
\begin{eqnarray}
{\rm tr} \int_{k}   \ln {\rm det}{\cal M} &=& \int_{k}\ln 
\bigl[k_{0}-E_{+}\bigr]\bigl[k_{0}+E_{+}\bigr]
\bigl[k_{0}-E_{-}\bigr]\bigl[k_{0}+E_{-}\bigr].
\end{eqnarray}
The eigenvalues appeared in $V_{eff}$ will be obtained in the following forms:
\begin{eqnarray}
E_{\pm}(\bmk) &=& \sqrt{\bigl( \sqrt{k^{2}_{\perp}+[\sigma+m+u\cos k_{3}]^{2}} \mp [ \mu + t\cos k_{3} ] \bigr)^{2} + |\Delta|^{2} }.
\end{eqnarray}

\vspace{2mm}

We employ the finite-temperature Matsubara formalism 
for our convenience of $k_{0}$-integration~[27-29]. 
The zero-temperature theory is converted into the Matsubara formalism by the following substitution in our theory:
\begin{eqnarray}
& & \int\frac{dk_{0}}{2\pi i} \to \sum_{n}\frac{1}{\beta}, \quad k_{0} \to i\omega_{n}, \quad \beta \equiv \frac{1}{T}, \quad ( T; {\rm temperature} ), \quad \omega_{n} \equiv \frac{(2n+1)\pi}{\beta}, \quad ( n=0,\pm 1,\pm 2, \cdots ),
\end{eqnarray}
After performing the discrete frequency summation, we get
\begin{eqnarray}
V_{eff} &=& \frac{\sigma^{2}}{2G} + \frac{|\Delta|^{2}}{2G}  
 -  \int^{\Lambda}\frac{d^{2}k_{\perp}}{(2\pi)^{2}}\int^{\pi}_{-\pi}\frac{dk_{3}}{2\pi}\Bigl( E_{+}+E_{-} +\frac{2}{\beta}\ln(1+e^{-\beta E_{+}})(1+e^{-\beta E_{-}}) \Bigr),   
\end{eqnarray}
where, $\Lambda$ is a two-dimensional momentum cutoff:
\begin{eqnarray}
\int^{\Lambda}\frac{d^{2}k_{\perp}}{(2\pi)^{2}}  &\equiv& \frac{1}{2\pi}\int^{\Lambda}_{0}|k_{\perp}|d|k_{\perp}|.
\end{eqnarray}
The gap equations, namely the stationary condition of $V_{eff}$ will be obtained as follows:
\begin{eqnarray}
0 &=& \frac{\partial V_{eff}}{\partial\sigma}  
= \frac{\sigma}{G} - \int^{\Lambda}  \frac{d^{2}k_{\perp}}{(2\pi)^{2}}\int^{\pi}_{-\pi}\frac{dk_{3}}{2\pi}\Bigg[ 
\frac{\partial E_{+}}{\partial\sigma}\tanh\frac{\beta}{2}E_{+} + \frac{\partial E_{-}}{\partial\sigma}\tanh\frac{\beta}{2}E_{-}  
\Bigg]
\end{eqnarray}
and
\begin{eqnarray}
0 &=& \frac{\partial V_{eff}}{\partial\overline{\Delta}}   
= \frac{\Delta}{2G}  - \int^{\Lambda} \frac{d^{2}k_{\perp}}{(2\pi)^{2}}\int^{\pi}_{-\pi}\frac{dk_{3}}{2\pi}\Bigg[ 
\frac{\partial E_{+}}{\partial\overline{\Delta}}\tanh\frac{\beta}{2}E_{+} +
\frac{\partial E_{-}}{\partial\overline{\Delta}}\tanh\frac{\beta}{2}E_{-} \Bigg].
\end{eqnarray}
Here, several derivatives  in the gap equations become
\begin{eqnarray}
\frac{\partial E_{\pm}}{\partial\sigma} &=& 
\frac{\sqrt{k^{2}_{\perp}+[\sigma+m+u\cos k_{3}]^{2}} \mp [\mu+t\cos k_{3}]}{\sqrt{\bigl( \sqrt{k^{2}_{\perp}+[\sigma+m+u\cos k_{3}]^{2}} \mp [\mu+t\cos k_{3}] \bigr)^{2} + |\Delta|^{2}}}
\frac{\sigma+m+u\cos k_{3}}{\sqrt{k^{2}_{\perp}+(\sigma+m+u\cos k_{3})^{2}}}
\end{eqnarray}
and
\begin{eqnarray}
\frac{\partial E_{\pm}}{\partial\overline{\Delta}} 
&=& \frac{1}{2}\frac{\Delta}{\sqrt{\bigl( \sqrt{k^{2}_{\perp}+[\sigma+m+u\cos k_{3}]^{2}} \mp [\mu+t\cos k_{3}] \bigr)^{2} + |\Delta|^{2}}}.   
\end{eqnarray}

\vspace{2mm}

Our gap equations have model parameters as $G$ ( coupling constant ), $\Lambda$ ( two-dimensional momentum cutoff ), $m$ ( gap parameter at the conical intersection ), 
$u$ ( a mass parameter alternating to the inter-layer direction ), $t$ ( transfer matrix between two nearest neighbor layers ) and $\mu$ ( a chemical potential ).
We choose the unit $\Lambda=1$ throughout this paper. The coupling constant $G$ will be chosen to satisfy the condition $\sigma,\Delta\ll \Lambda$.
In this paper, we do not consider a coexistent phase of $\sigma\ne 0$ and $\Delta\ne 0$,
and study phases of both $\sigma\ne 0$, $\Delta=0$ and $\sigma=0$, $\Delta\ne 0$ separately.

\vspace{2mm}

First, we investigate the critical coupling for dynamical generation of $\sigma$.
At $m=u=\Delta=T=0$, the self-consistent gap equation for $\sigma$ becomes like
\begin{eqnarray}
0 &=& \frac{\sigma}{G} -\frac{1}{2\pi}\int^{\Lambda}_{0}k_{\perp}dk_{\perp}\frac{\sigma}{\sqrt{k^{2}_{\perp}+\sigma^{2}}}
\end{eqnarray}
and this equation gives the critical coupling $G_{cr}$ as
\begin{eqnarray}
G_{cr} &=& \frac{2\pi}{\Lambda}.
\end{eqnarray}
Hence, one cannot find any effect of $t$ in $G_{cr}$ at $m=u=\Delta=T=0$, namely the inter-layer hopping effect of $t$ ( and also $\mu$ ) disappears in $G_{cr}$
at this case.
If we consider the case $\Delta\ne 0$ with $m=u=T=0$, $G_{cr}$ will obtain an expression which includes an effect of $t$,
though we have to solve the gap equations of $\sigma$ and $\Delta$ simultaneously for our confirmation of coexistence of
dynamically generated $\sigma$ and $\Delta$.
This is an interesting problem, though it is beyond scope of this paper.

\vspace{2mm}

Numerical results from the gap equation for $\Delta$ are given in Figs. 1, 2 and 3.
Figure 1 shows the BCS gap function $\Delta$ at $\sigma=0$ as a function of the inter-layer transfer parameter $t$. We set parameters as
$\Lambda=1$ (unit), $G=4$, $T=u=\mu=0$ 
( $\mu=0$ corresponds to the "Dirac" vacuum state ). 
The four examples of $m=0,0.1,0.2,0.3$ are depicted in this figure.
When $t$ ( $m$ ) increases, $\Delta$ increases ( decreases ).
There is no non-trivial solution for $\Delta$ at the cases of $t=0,m=0.3$ and $t=0.1,m=0.3$,
and this fact indicates the existence of a "critical transfer" ( and  also, a "critical coupling" ) for dynamical generation of finite $\Delta$ 
in our theory.

\vspace{2mm}

Figure 2 gives the BCS gap function $\Delta$ at $\sigma=0$ as a function of $t$. 
The parameters $\Lambda$, $G$, $T$ and $u$ are set as same as Fig. 1, while we choose $\mu=0.2$. 
Increasing of $\mu$ causes the enhancement of $\Delta$.

\vspace{2mm}

Figure 3 shows $\Delta$ at $\sigma=0$ shown as a function of $t$. 
The parameters $\Lambda$, $G$, $T$ and $u$ are set as same as Figs. 1 and 2, 
and chemical potential is chosen as $\mu=0.4$. 
In this figure, $\Delta$ decreases when $t$ increases.
Hence, $t$-dependence of $\Delta$ is not trivial, and it is affected by $\mu$.
It is also the case that, when $m$ is small enough, the difference of choices of numerical values of $\mu$ disappears at large $t$
( effect of $\mu$ is washed out by a large $t$ ).
See the differences of energy scales of Figs. 1, 2 and 3.

\vspace{2mm}

Temperatre dependence of gap function $\Delta$ is shown in Figs. 4 and 5. 
We choose three values of $\mu$ in Fig. 4, while the three examples of $t$ are chosen in Fig. 5.
In both of these figures, the ratio $\Delta(T=0)/T_{c}$ ( $T_{c}$; critical temperature ) becomes $1.7\pm0.1$,
i.e., the BCS universal constant $\Delta(T=0)/T_{c}=1.76$ is almost satisfied, 
though it seems the case that there are tiny but finite deviations from the constant.
The gap function always show the character of second order phase transition.

\vspace{2mm}

We summarize our conclusions as the following tendencies:
(1) When $m$ increases, $\Delta$ decreases.
(2) When $\mu$ increases, $\Delta$ increases.
(3) $\Delta$ depends on $t$ non-trivially, and its dependence on $t$ is qualitatively changed by choices of numerical values of $\mu$.

\vspace{2mm}

If we consider a real substance like graphite, $\mu$ corresponds to a charge doping concentration to the system.
Our results indicate that a charge doping and an inter-layer coupling do not always give cooperative effects to an amplitude of a BCS gap function. 
The results of this paper can also be used to consider superconductivity of ${\rm MgB_{2}}$~[30]
which also has the honeycomb lattice symmetry~[7].
It is very interesting for us to use parameters obtained by an ${\it ab}$ ${\it initio}$ band structure 
calculation or a molecular orbital calculation for getting numerical values of $t$, so forth.

\begin{figure}

\caption{$\Delta$ shown as a function of $t$. 
We set parameters as $\Lambda=1$ (unit), $G=4$, $T=u=\sigma=\mu=0$
( $\mu=0$ corresponds to the "Dirac" vacuum state ). 
The four examples of $m=0,0.1,0.2,0.3$ are depicted.}

\caption{$\Delta$ shown as a function of $t$. 
We set parameters as $\Lambda=1$ (unit), $G=4$, $T=u=\sigma=0$ and $\mu=0.2$. 
The four examples of $m=0,0.1,0.2$ are depicted in this figure.}

\caption{$\Delta$ shown as a function of $t$. 
We set parameters as $\Lambda=1$ (unit), $G=4$, $T=u=\sigma=0$ and $\mu=0.4$. 
The four examples of $m=0,0.1,0.2,0.3$ are depicted in this figure.}

\caption{$\Delta$ as a function of  temperature $T$. 
We have set parameters as $\Lambda=1$ ( unit ), $G=4$, $u=\sigma=0$, $m=0.1$ and $t=0.2$.
Three examples of $\mu=0,0.2,0.4$ are shown in this figure.}

\caption{$\Delta$ as a function of  temperature $T$. 
We have set parameters as $\Lambda=1$ ( unit ), $G=4$, $u=\sigma=0$, $m=0.1$ and $\mu=0.2$.
Three examples of $t=0,0.2,0.4$ are shown in this figure.}

\end{figure}

\end{document}